\documentclass[11pt,a4paper]{article}
\usepackage{mcite}
\usepackage{epsfig}

\epsfverbosetrue

\def\3{\ss}

\newcommand{\tev}{{\rm Te}\kern-1.pt{\rm V}}
\newcommand{\gev}{{\rm Ge}\kern-1.pt{\rm V}}
\newcommand{\mev}{{\rm Me}\kern-1.pt{\rm V}}
\newcommand{\kev}{{\rm Ke}\kern-1.pt{\rm V}}
\newcommand{\gevsq}{\mbox{$\mathrm{{\rm Ge}\kern-1.pt{\rm V}}^2$}}
\newcommand{\gevmsq}{\mbox{$\mathrm{{\rm Ge}\kern-1.pt{\rm V}}^{-2}$}}

\newcommand{\BR}        {\mbox{$\mathcal{B}$}}

\newcommand{\menor} {\mbox{\raisebox{-0.4ex}
{$\;\stackrel{<}{\scriptstyle \sim}\;$}}}

\newcommand{\sla}[1]{/\!\!\!#1}

\begin{document}
\begin{titlepage}

\begin{center}
\begin{huge}
\bf Feasibility of Searches for a Higgs Boson using \boldmath{$H\rightarrow W^+W^-\rightarrow l^+l^-\sla{p_T}$} and High $P_T$ Jets at the Tevatron  \\
\end{huge}

\vspace{2.cm}

\Large Bruce Mellado, William Quayle and Sau Lan Wu\\
\vspace{0.5cm}
{\Large\it Physics Department \\
University of Wisconsin - Madison \\
   Madison, Wisconsin 53706 USA }

\vspace{1.5cm}

\begin{abstract}
\noindent  The sensitivity of Standard Model Higgs boson searches at the Tevatron experiments with a mass $135<M_H<190\,\gev/c^2$  using the  channel $H\rightarrow W^+W^-\rightarrow l^+l^-\sla{p_T} (l=e,\mu)$  is discussed. Three new event selections involving Higgs in association with one or two high $P_T$ hadronic jets are discussed. Using Leading Order Matrix Elements and a conservative cut-based analysis a $95\,\%$ confidence level exclusion on $\sigma\times \BR (H\rightarrow W^+W^-)$, 1.6 times larger than that predicted by  the Standard Model for $M_H=165\,\gev/c^2$, may be achieved with $5\,$fb$^{-1}$ of integrated luminosity. By combining these three event selections with the existing analysis, the sensitivity of CDF and D0 could improve significantly.
\end{abstract}
\end{center}
\setcounter{page}{0}
\thispagestyle{empty}

\end{titlepage}

\newpage

\pagenumbering{arabic}

\section{Introduction}
\label{sec:introduction}

In the Standard Model (SM) of electro-weak and strong
interactions, there are four types of  gauge vector bosons (gluon,
photon, $W^\pm$ and $Z$) and twelve types of fermions (six quarks and six
leptons)~\cite{np_22_579,prl_19_1264,sal_1968_bis,pr_2_1285}.
These particles have been observed experimentally. At present, all
the data obtained from the many experiments in particle physics
are in agreement with the Standard Model.  In the Standard Model
there is one particle, the Higgs boson, that is responsible for
giving masses to all of the other 
particles~\cite{prl_13_321,pl_12_132,prl_13_508,pr_145_1156,prl_13_585,pr_155_1554}.
In this sense, the Higgs particle occupies a unique position.
  
Before the startup of the Large Hadron Collider (LHC) the Tevatron remains the high energy frontier. Higgs searches at the Tevatron are a priority for high energy physics. The Standard Model (SM) Higgs may be produced at the Tevatron via several
mechanisms. The  Higgs is expected to be produced predominantly via
gluon-gluon fusion~\cite{prl_40_11_692} and Higgs strahlung off a $Z$ or $W^\pm$ boson. The third dominant process at the Tevatron is vector boson
fusion (VBF) for $M_H\menor 200\,\gev/c^2$~\cite{pl_136_196,pl_148_367} . A significant fraction of the Higgs produced via these mechanisms will be associated with at least one high transverse momentum ($P_T$) hadronic jet. The kinematics of Higgs signals in association with jets differ significantly from that of known SM backgrounds. These properties can be exploited to select corners of the phase space where the expected signal-to-background ratios are larger than in a purely inclusive approach. 

\begin{figure}[h]
{\centerline{\epsfig{figure=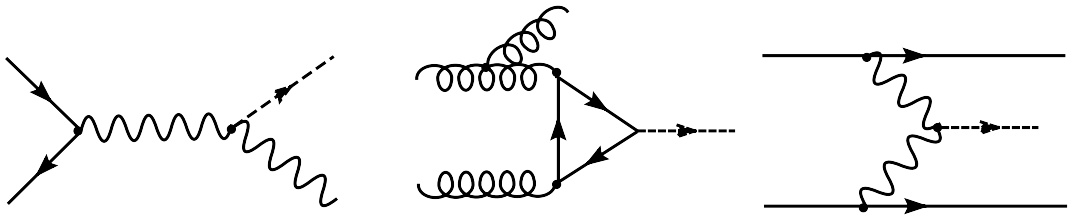,width=14cm}}}
\caption[]{Leading diagrams for the Higgs production in association with jets at the Tevatron: $VH, V=Z,W^\pm (V\rightarrow q\overline{q})$ (left), Gluon-Gluon fusion (center) and Vector Boson Fusion (right).}
 \label{fig:higgs}
\end{figure}

Searching for a SM Higgs boson at the Tevatron via its decay $H \rightarrow W^+W^-$ was first considered in~\cite{prl_82_25,pr_59_093001}.  The relevance of observing a low mass SM Higgs in association with one or two jets of high $P_T$ at the LHC has been pointed out by a number of authors~\cite{JHEP_12_5,pr_60_113004,pr_59_014037,pr_61_093005,pl_431_410,pl_611_60}.  The ATLAS and CMS experiments have confirmed these expectations with detailed detector simulations~\cite{EurPhysJ_32_19,CMSPTDR}. The feasibility of the Higgs searches using the decay $H\rightarrow W^+W^-\rightarrow l^+l^- (l=e,\mu)$ in association with one high $P_T$ jet is not discussed in the literature and  is investigated here. 

In this paper we study the sensitivity of the Tevatron experiments to a SM Higgs with a mass $135<M_H<190\,\gev/c^2$ using the decay $H\rightarrow W^+W^-\rightarrow l^+l^- (l=e,\mu)$ in association with hadronic jets. We evaluate the feasibility of three event selection schemes involving events with one or two tagging jets and the impact of these selections on the sensitivity to SM Higgs searches for the D0 and CDF experiments. In addition to the potential enhancement of the sensitivity, a study at the Tevatron of the final states discussed here will bring significant benefit to future searches at the LHC in terms of understanding QCD effects on the production of known SM particles in these corners of the phase space. 
		
\section{MC Generation of Relevant Processes}
\label{sec:mc}

The final state under study consists of the following signatures
\begin{itemize}
\item Two high $P_T$ leptons ($e,\mu$) of opposite sign.
\item  Missing transverse momentum ($\sla{p_T}$) inconsistent with detector fluctuations.
\item One or two high $P_T$ hadronic jets.
\end{itemize}

For the generation of the Higgs signal processes specified in Section~\ref{sec:introduction}, Leading Order (LO) Matrix Elements (ME) were interfaced with the Pythia~\cite{cpc_82_74,cpc_135_238} and Herwig~\cite{JHEP_0101_010} packages to provide  parton showers and hadronization effects. The cross-section computation and the event generation were performed using the parameterization of the parton density functions provided by CTEQ6~\cite{JHEP_0207_012}.  QCD Next-to-Leading-Order (NLO) effects are known to be sizeable for the $gg\rightarrow Hj$ process. However, QCD NLO effects are not known for all of the backgrounds. Therefore, the potential improvement in sensitivity due to these effects were not considered here.

\begin{table}[ht]
\begin{center}
\begin{tabular}{||c|c|c||c|c|c||}
\hline 

$gg\rightarrow Hj$ & $VBF H$ & $VH$ &    $WW+ 1j$ & $t\overline{t}$  &     $Z\rightarrow\tau^+\tau^-+1j$              
\\ \hline
 96.5 &   37.6  &       61.3 &      3480 &    7200   & 55200 \\ \hline
\end{tabular}
\caption{Cross-sections (in fb) of the SM Higgs signal ($M_H=165\,\gev/c^2$) and the main background processes considered in this paper (see text).}
\label{tab:crosssec}
\end{center}
\end{table}

The cross-sections and kinematics of the Higgs processes were cross-checked with the corresponding ME provided by MCFM~\cite{MCFM} and ALPGEN~\cite{JHEP_0307_001}. Table~\ref{tab:crosssec} displays the cross-sections for Higgs signal and the main background processes. The cross-sections reported in Table~\ref{tab:crosssec} do not take into account the branching fraction of $Z/W^\pm$ (except for $Z\rightarrow\tau^+\tau^-$) or any of its decay products. The cross-section of Higgs via gluon-gluon fusion,\footnote{For the computation of the Higgs production cross-section via gluon-gluon fusion the top mass was assumed $M_{Top}=174.2\,\gev/c^2$.} of  $WW$ production\footnote{Matrix elements involved in $p\overline{p}\rightarrow W^{\pm}W^{\pm}jj$ are included. The contribution from these diagrams is expected to be negligible due to the requirement that the two charged leptons be of opposite sign (see Section~\ref{sec:preselection}). Diagrams involving two gluons in the final state have not been taken into account.} and of $Z\rightarrow\tau^+\tau^-$\footnote{The cross-section for $Z\rightarrow\tau^+\tau^-+1j$ was calculated for $40<M_{\tau\tau}<200\,\gev/c^2$. }  in association with one jet quoted in Table~\ref{tab:crosssec} is defined for jet $P_T>10\,\gev/c^2$ and $\left|\eta\right|<100$. The $WW$ process was generated with ALPGEN. The cross-sections have been evaluated by setting the renormalization and factorization scales to $\sqrt{M^2+\sum P_T^2}$ where $M$ is the mass of the weak bosons and the $\sum P_T^2$ stands for the scalar sum of partons in the final state. The cross-section for $t\overline{t}$ production used here corresponds to the measured value~\cite{CDF8148} and the generation was performed with MC@NLO~\cite{JHEP_0206_029,JHEP_0308_007}.

The contribution from processes in which at least one lepton arises from a jet faking a lepton  is normalized with respect to the effective cross-section of $WW+jets$. The ratio of the cross-section of fakes to $WW+jets$ is assumed to be the same as in~\cite{CDFWWWinter07}. The contribution from $Z\rightarrow l^+l^-$ with $l=e,\mu$ is expected to be negligible due to the requirement of a minumum $\sla{p_T}$, a requirement of the minimum transverse momentum of the lepton system and an upper bound on the  invariant mass of the two leptons (see Section~\ref{sec:selection}).\footnote{The contribution from $Z\rightarrow l^+l^-+jets$ with $l=e,\mu$ was evaluated with Pythia using the $2 \rightarrow 2$ ME.} The contribution from  $Z\rightarrow \tau^+\tau^-, ZZ, ZW^\pm$ and $W^\pm\gamma$ are important. For the sake of simplicity, we assume that the survival probability of processes involving two gauge bosons $ZZ, ZW^\pm, W^\pm\gamma$ against cuts on jets presented in Section~\ref{sec:selection} are the same as for $WW$. The contribution from $Z\rightarrow \tau^+\tau^-+jets$ was modeled with Pythia using the $2\rightarrow 2$ ME. 

In order to emulate detector effects a simple fast simulation program was implemented. Hadrons are clusterized using a classical cone algorithm with a $\Delta R<0.4.$\footnote{$\Delta R$ is defined as $\sqrt{ \left(\Delta\eta\right)^2+\left(\Delta\phi\right)^2}$, where $\eta=-\ln{\tan{{\theta \over 2}}}$ and $\theta$ and $\phi$ are the polar and azimuthal angles, respectively.} The energy of the resulting hadronic jets was smeared according to a resolution function of the form ${\sigma E \over E }= {a \over \sqrt{E} }\oplus b $. The values of $a$ are set to $0.5 ,0.8$ for $\left|\eta\right|<0.9$ and $0.9<\left|\eta\right|<3$, respectively. The values of $b$ are set to $0.03 ,0.05$ for $\left|\eta\right|<0.9$ and $0.9<\left|\eta\right|<3$, respectively. The resolution function for electro-magnetic depositions was set to ${\sigma E \over E }= {0.15 \over \sqrt{E} }\oplus 0.02$. The reconstructed $\sla{p_T}$  follows the following resolution function $\sigma\left(\sla{p_{x(y)}}\right)=0.6 \sqrt{\sum E_T}$ where $\sla{p_{x(y)}}$ is the $x$ ($y$) component of  $\sla{p_T}$ and  $\sum E_T$ is the scalar sum of the transverse energy particles within $\left|\eta\right|<3.5$.

Electron and muon identification efficiencies are assumed to be $0.9$ in the range $\left|\eta\right|<1.5$~\cite{Herndon07}. In the event selection presented in Section~\ref{sec:selection3}, b-tagging capabilities are used. It is assumed that the b-tagging efficiency is 0.5 for $\left|\eta\right|<1$. In the forward region (1$<\left|\eta\right|<1.9$) the b-tagging efficiency is parameterized with a linear function $1.05-0.55\left|\eta\right|$~\cite{Herndon07}.

\section{Event Selection}
\label{sec:selection}

In this Section we present the results of three event selections in different corners of the phase space. The three analyses proposed here are orthogonal to one another and relatively good signal-to-background ratios may be achieved. The three event selections presented in Sections~\ref{sec:selection1}-\ref{sec:selection3} exploit the particular kinematics of the jets in the events produced by the three main signal processes referred to in Section~\ref{sec:introduction}.

\subsection{Preselection}
\label{sec:preselection}

The final event selections are preceded by a preselection after which the backgrounds are expected to be dominated by $WW+jets$ and $t\overline{t}$ production. The contribution from backgrounds in which one or two leptons arise from misidentification of jets or $b\overline{b}$ events is not expected to be large.

\begin{table}[ht]
\begin{center}
\begin{tabular}{||c||c|c|c||c|c|c||}
\hline 
Cut  & $gg\rightarrow Hj$ & $VBF H$ & $VH$ &                 $WW+1j$ & $t\overline{t}$   & $Z\rightarrow\tau^+\tau^-+ 1j$
\\ \hline
{\bf a} &       2.50 &       0.97 &       2.73 &     175.95 &     206.02 &     143.22
\\ \hline
 {\bf b} &       2.37 &       0.92 &       2.24 &     143.23 &     190.49 &      56.55
\\ \hline
 {\bf c} &       1.89 &       0.73 &       1.36 &      69.71 &      49.13 &      54.75
\\ \hline
 {\bf d} &       1.68 &       0.64 &       1.14 &      49.96 &      34.84 &       8.19
\\ \hline
\end{tabular}
\caption{Effective cross-sections (in fb) for signal ($M_H=165\,\gev/c^2$) and main background processes after pre-selection cuts specified in Section~\ref{sec:preselection}.}
\label{tab:preselection}
\end{center}
\end{table}

The cuts applied in the pre-selection are the following:
\begin{itemize}
\item[\bf a)] Two opposite sign leptons ($e,\mu$) in $\left|\eta\right|<1.5$ with $P_T>20\,\gev/c$ for the leading lepton and $P_T>10\,\gev/c$ for the sub-leading one. Veto events with a third lepton in $\left|\eta\right|<1.5$ with $P_T>10\,\gev/c$.
\item[\bf b)] Presence of missing transverse momentum of at least $20\,\gev/c$.
\item[\bf c)] Requirements on the invariant mass of the leptons, $20<M_{ll}<70\,\gev/c^2$.
\item[\bf d)] Lepton azimuthal angle difference, $\Delta\phi_{ll}<2.5\,$rad and transverse momentum of the leptonic system, $P_{T ll}>35\,\gev/c$.
\end{itemize}

Table~\ref{tab:preselection} displays the effective cross-sections for the three signal processes and backgrounds for cuts {\bf a-c}. It is relevant to note that no requirement on the presence of jets has been made in Table~\ref{tab:preselection}. The requirements on the minimum $\sla{p_T}$ and the invariant mass of the two leptons in cuts {\bf c} and {\bf d} enhances the ratio of signal to the main backgrounds by about a factor of two. These cuts diminish the discriminating power of variables such as the transverse mass of the leptons and the $\sla{p_T}$  or the azimuthal angle difference between the two leptons.\footnote{Here we considered a simple-minded cut-based analysis. A more sophisticated multivariate analysis could take advantage any residual discrimination remaining in these variables.} The requirements applied in cut {\bf d} are mainly intended to suppress the Drell-Yan process $Z\rightarrow\tau^+\tau^-\rightarrow l^+l^-\sla{p_T}$. The presense of $\sla{p_T}$ is a strong discrimantor against this process. However, in events with jets studied here a cut of $\sla{p_T}$ is not enough to achieve the neccesary rejection. The requirement of the minimum transverse momentum of the lepton pair is particularly effective in reducing Drell Yan backgrounds. 

\subsection{ Selection I}
\label{sec:selection1}

\begin{figure}[t]
{\centerline{\epsfig{figure=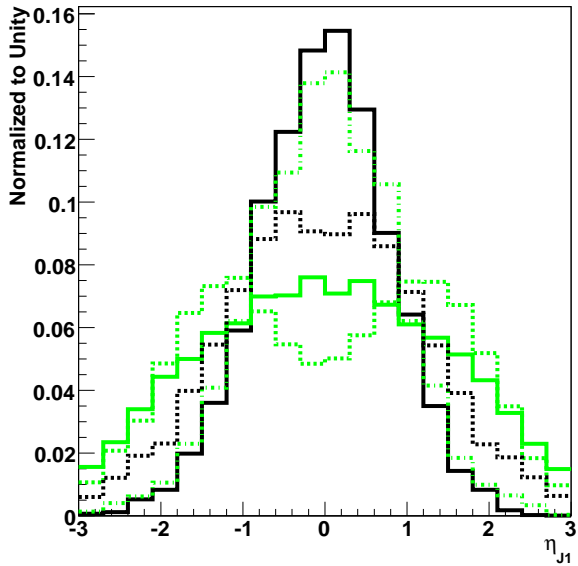,width=8cm}}}
\caption[]{The pseudorapidity of the leading jet for signal and background processes after the application of preselection cuts presented in Section~\ref{sec:preselection}. The solid, dashed and dotted dashed light colored histograms correspond to gluon-gluon fusion,  VBF and $VH$ signal production, respectively. The solid and dashed black histograms correspond to $t\overline{t}$ and $WW+jets$, respectively.  Histograms are normalized to unity.} \label{fig:selection1}
\end{figure}

\begin{table}[ht]
\begin{center}
\begin{tabular}{||c||c|c|c||c|c|c|c||c||}
\hline 
Cut  & $gg\rightarrow Hj$ & $VBF H$ & $VH$ &                 $WW$ & $t\overline{t}$                                 & $Z\rightarrow\tau^+\tau^-$                         & Other & S/B 
\\ \hline
 {\bf Ia} &       0.95 &       0.60 &       0.97 &      11.51 &      34.52 &       8.14 &       6.26 &       0.04
\\ \hline
 {\bf Ib} &       0.85 &       0.46 &       0.52 &      10.82 &      25.96 &       7.50 &       5.89 &       0.04
\\ \hline
 {\bf Ic}&       0.74 &       0.34 &       0.36 &       9.57 &       9.28 &       6.25 &       5.25 &       0.05
\\ \hline
  {\bf Id}&       0.33 &       0.16 &       0.07 &       2.86 &       1.17 &       1.05 &       1.61 &       0.08
\\ \hline
$\left|\eta_j\right|>1.5$ &       0.26 &       0.12 &       0.04 &       1.99 &       0.71 &       0.55 &       1.14 &       0.10
\\ \hline
$\left|\eta_j\right|>1.75$ &       0.19 &       0.08 &       0.02 &       1.33 &       0.36 &       0.28 &       0.78 &       0.11
\\ \hline
\end{tabular}
\caption{Effective cross-sections (in fb) for signal and background processes after selection cuts specified in Section~\ref{sec:selection1}.}
\label{tab:selection1}
\end{center}
\end{table}

The selection proposed here takes advantage of the fact that the leading jet produced in association with Higgs via the gluon-gluon fusion and the VBF processes tend to be more forward than in background events.  This is illustrated in Figure~\ref{fig:selection1}. Figure~\ref{fig:selection1} displays the pseudorapidity distribution of the leading jet in the event for signal and background processes, after the application of the preselection requirements given in Section~\ref{sec:preselection}. 

Due to the potential contribution from Higgs signal produced via VBF it is important to reconstruct jets to the very forward region. Therefore in this selection we assume the ability to reconstruct jets of $P_T>15\,\gev/c$ in the range $\left|\eta\right|<3$. 

The event selection is comprised of the following cuts:
\begin{itemize}
\item [\bf Ia.] At least one jet with $P_T>15\,\gev/c$ in the range $\left|\eta\right|<3$.
\item [\bf Ib.] If the event has at least two jets with $P_T>15\,\gev/c$ in the range $\left|\eta\right|<3$ it is required that the event fails the two-jet event selections presented in Sections~\ref{sec:selection2}-\ref{sec:selection3}.\footnote{In Selection II the cut $\Delta\eta_{jj}>3$ is used (see Section~\ref{sec:selection2}).} 
\item [\bf Ic.] If a second jet is found with $P_T>15\,\gev/c$ and $\left|\eta\right|<1.25$ the event is rejected. 
\item [\bf Id.] Require that the leading jet be relatively forward, $1.25<\left|\eta\right|<3$.
\end{itemize}

Table~\ref{tab:selection1} displays the effective cross-section for signal and background processes after the application of the preselection cuts in Section~\ref{sec:preselection} and the above cuts. Cut {\bf Ib} is introduced to avoid double counting of events  passing selections presented in Sections~\ref{sec:selection2}-\ref{sec:selection3}.  Cut ${\bf Ic}$ is introduced instead of a full jet veto in the entire range $\left|\eta\right|<3$ in order to enhance the contribution from Higgs signal events produced by the VBF mechanism. These types of events evolve from a residual fraction of events that do not pass the stringent two-jet event selection presented in Section~\ref{sec:selection2}, specifically the requirement that the pseudorapidity difference between the jets be greater than a large value. A significant fraction of Higgs events produced via the $VH$ mechanism are lost after cut {\bf Ic}. In order to recover these events a third event selection is presented here (see Section~\ref{sec:selection3}). 

Due to  the application of cuts {\bf Ic}-{\bf Id} the two leading jets in the event tend to be in a pseudorapidity range in which the b-tagging efficiency is rather small. Therefore the b-tagging capability was not used to further suppress $t\overline{t}$ background. 

The rows after cut {\bf Id} in Table~\ref{tab:selection1} correspond to  tighter cuts on the pseudorapidity of the leading jets. By tightening the requirement on the "forwardness" of the leading  jet the signal-to-background ratio improves significantly with respect to cut {\bf Id}.  As far as the sensitivity is concerned, the optimal value of the range of the pseudorapidity of the leading jet is $1.25<\left|\eta\right|<3$. 

\subsection{ Selection II}
\label{sec:selection2}

\begin{figure}[t]
{\centerline{\epsfig{figure=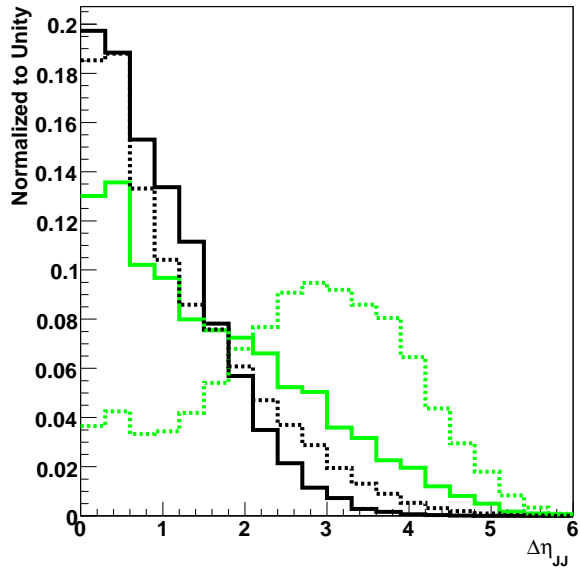,width=8cm}}}
\caption[]{The pseudorapidity difference of the two leading jets for signal and background processes after the application of preselection cuts presented in Section~\ref{sec:preselection}. The solid, dashed light colored histograms correspond to gluon-gluon fusion, VBF signal production, respectively. The solid and dashed black histograms correspond to $t\overline{t}$ and $WW+jets$, respectively.  Histograms are normalized to unity. } \label{fig:selection2}
\end{figure}

\begin{table}[ht]
\begin{center}
\begin{tabular}{||c||c|c|c||c|c|c|c||c||}
\hline 
Cut  & $gg\rightarrow Hj$ & $VBF H$ & $VH$ &                 $WW$ & $t\overline{t}$       & $Z\rightarrow\tau^+\tau^-$                           & Other & S/B 
\\ \hline
 {\bf IIa} &       0.28 &       0.41 &       0.66 &       2.50 &      29.62 &       2.66 &       1.31 &       0.04
\\ \hline
  {\bf IIb}&       0.06 &       0.24 &       0.01 &       0.27 &       1.14 &       0.34 &       0.14 &       0.16
\\ \hline
$\Delta\eta_{jj}>3.0$ &       0.04 &       0.17 &       0.00 &       0.13 &       0.39 &       0.17 &       0.07 &       0.29
\\ \hline
$\Delta\eta_{jj}>3.5$ &       0.02 &       0.11 &       0.00 &       0.06 &       0.11 &       0.07 &       0.04 &       0.49
\\ \hline
$\Delta\eta_{jj}>4.0$ &       0.01 &       0.06 &       0.00 &       0.03 &       0.03 &       0.02 &       0.01 &       0.80
\\ \hline
\end{tabular}
\caption{Effective cross-sections (in fb) for signal and background processes after selection cuts specified in Section~\ref{sec:selection2}.}
\label{tab:selection2}
\end{center}
\end{table}

The event selection presented here is tuned to enhance the Higgs contribution from the VBF mechanism. This is illustrated in Figure~\ref{fig:selection2}, in which the shapes of the distributions are compared for signal and background processes. It is important to note that the distribution for the VBF signal process shown in Figure~\ref{fig:selection2} enhances the fraction of the events with a small angular separation with respect to the one predicted by the fixed order NLO computation by MCFM. The corresponding distribution for the signal process via gluon-gluon fusion, although it is expected to be small,  is not reliable since in the MC generation used here no ME correction was applied on the sub-leading jet. 

The rate and angular distributions of additional jet activity with $P_T>15\,\gev/c$ was investigated in signal and background processes. This is motivated by the fact that the leading signal contribution comes from a color singlet exchange and a reduced rate of hadronic jets is expected in the signal-like region. A  veto on additional jet activity was found to give additional discriminating power against $t\overline{t}$ production, but was not used here. Further investigation could be performed by lowering the $P_T$ threshold to $10\,\gev/c$.

The event selection is comprises the following cuts:
\begin{itemize}
\item [\bf IIa.] At least two jets with $P_T>15\,\gev/c$ in the range $\left|\eta\right|<3$.
\item [\bf IIb.] Large difference in pseudorapidity between the two leading jets, $\Delta\eta_{jj}>2.5$. 
\end{itemize}

Table~\ref{tab:selection2} displays the effective cross-sections for signal as well as backgrounds and the signal to background ratios after the application of cuts {\bf IIa}-{\bf IIb}.  The last four rows in Table~\ref{tab:selection2} correspond to the application of tighter cuts on $\Delta\eta_{jj}$. The optimal value on the cut on the di-jet pseudorapidity difference to achieve the best $95\,\%$ confidence limit is $\Delta\eta_{jj}>3$.

\subsection{ Selection III}
\label{sec:selection3}

\begin{figure}[t]
{\centerline{\epsfig{figure=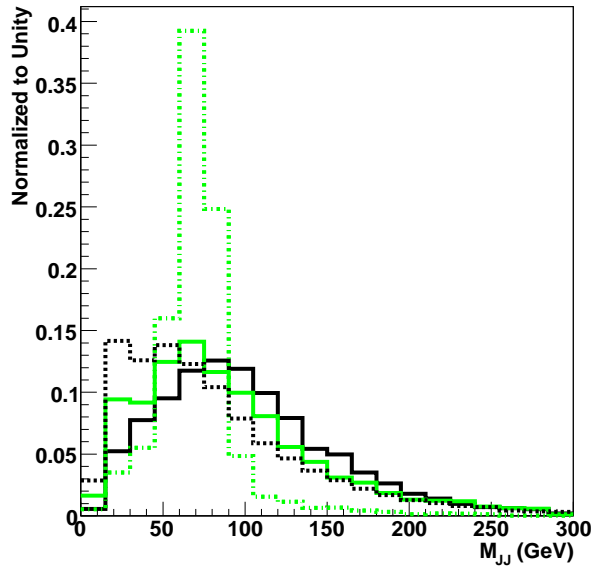,width=8cm} }  }
\caption[]{The invariant mass of the two leading jets after the application of preselection cuts presented in Section~\ref{sec:preselection}.  The solid, dashed and dotted dashed light colored histograms correspond to gluon-gluon fusion, VBF and $VH$ signal production, respectively. The solid and dashed black histograms correspond to $t\overline{t}$ and $WW+jets$, respectively. Histograms are normalized to unity.} \label{fig:selection3}
\end{figure}

The event selection proposed here is intended to enhance the efficiency of tagging $VH$ with $V\rightarrow qq^{\prime}$. The event selection is comprised of the following cuts:
\begin{itemize}
\item [\bf IIIa.] At least two jets with $P_T>20\,\gev/c$ in the range $\left|\eta\right|<2.5$.
\item [\bf IIIb.] It is required that the event does not pass the selection presented in Sections~\ref{sec:selection2}. 
\item [\bf IIIc.] B-jet veto selection (see Section~\ref{sec:mc}).
\item [\bf IIId.] Invariant mass of the two leading jets, $50<M_{jj}<80\,\gev/c^2$.
\end{itemize}

After the application of cuts {\bf IIIa}-{\bf IIIc} the $t\overline{t}$ process becomes the dominant one. As the two leading jets in this selection are central, the ability to tag $b$-jets becomes essential. The enhancement  of $b$-tagging efficiency is, however, not the only handle  to further suppress the $t\overline{t}$ background.

\begin{table}[ht]
\begin{center}
\begin{tabular}{||c||c|c|c||c|c|c|c||c||}
\hline 
Cut  & $gg\rightarrow Hj$ & $VBF H$ & $VH$ &  $WW$ & $t\overline{t}$   & $Z\rightarrow\tau^+\tau^-$  & Other & S/B 
\\ \hline
 {\bf IIIa} &       0.16 &       0.30 &       0.53 &       1.54 &      26.79 &       1.89 &       0.80 &       0.03
\\ \hline
 {\bf IIIb} &       0.15 &       0.20 &       0.53 &       1.49 &      26.57 &       1.82 &       0.77 &       0.03
\\ \hline
 {\bf IIIc} &       0.14 &       0.20 &       0.51 &       1.47 &       9.94 &       1.80 &       0.76 &       0.06
\\ \hline
 {\bf IIId} &       0.05 &       0.02 &       0.37 &       0.43 &       2.68 &       0.39 &       0.23 &       0.12
\\ \hline
\end{tabular}
\caption{Effective cross-sections (in fb) for signal and background processes after selection cuts specified in Section~\ref{sec:selection3}.}
\label{tab:selection3}
\end{center}
\end{table}

\section{Results and Conclusions}
\label{sec:results}

\begin{table}[t]
\begin{center}
\begin{tabular}{||c||c|c|c||c|c||c|c|c||}
\hline 

& \multicolumn {3}{c||} {Analysis I}  & \multicolumn {2}{c||} {Analysis II}  & \multicolumn {3}{c||} {Analysis III}  \\ \hline
 $M_{H}$ & $gg\rightarrow Hj$ & $VBF H$ & $VH$ & $gg\rightarrow Hj$ & $VBF H$   & $gg\rightarrow Hj$ & $VBF H$ & $VH$
\\ \hline
135 &       0.16 &       0.06 &       0.05 &       0.02 &       0.08 &       0.02 &       0.01 &       0.22
\\ \hline
140 &       0.20 &       0.09 &       0.05 &       0.02 &       0.10 &       0.03 &       0.01 &       0.26
\\ \hline
145 &       0.23 &       0.10 &       0.06 &       0.03 &       0.11 &       0.03 &       0.02 &       0.29
\\ \hline
150 &       0.25 &       0.12 &       0.06 &       0.03 &       0.13 &       0.03 &       0.02 &       0.31
\\ \hline
155 &       0.28 &       0.14 &       0.07 &       0.03 &       0.14 &       0.04 &       0.02 &       0.32
\\ \hline
160 &       0.32 &       0.16 &       0.07 &       0.04 &       0.17 &       0.04 &       0.02 &       0.37
\\ \hline
165 &       0.33 &       0.16 &       0.07 &       0.04 &       0.17 &       0.05 &       0.02 &       0.37
\\ \hline
170 &       0.30 &       0.15 &       0.06 &       0.04 &       0.16 &       0.04 &       0.02 &       0.33
\\ \hline
175 &       0.26 &       0.14 &       0.05 &       0.03 &       0.15 &       0.04 &       0.02 &       0.29
\\ \hline
180 &       0.23 &       0.12 &       0.04 &       0.03 &       0.13 &       0.03 &       0.02 &       0.25
\\ \hline
185 &       0.18 &       0.10 &       0.03 &       0.02 &       0.10 &       0.03 &       0.01 &       0.19
\\ \hline
190 &       0.14 &       0.08 &       0.02 &       0.02 &       0.09 &       0.02 &       0.01 &       0.16
\\ \hline
195 &       0.12 &       0.07 &       0.02 &       0.02 &       0.07 &       0.02 &       0.01 &       0.13
\\ \hline
200 &       0.10 &       0.06 &       0.02 &       0.01 &       0.06 &       0.02 &       0.01 &       0.11
\\ \hline
\end{tabular}
\caption{Expected effective cross-sections (in fb) for the three signal processes and the three event selections considered in Section~\ref{sec:selection} as a function of the Higgs mass (in $\gev/c^2$).}
\label{tab:signals}
\end{center}
\end{table}

\begin{figure}[t]
{\centerline{\epsfig{figure=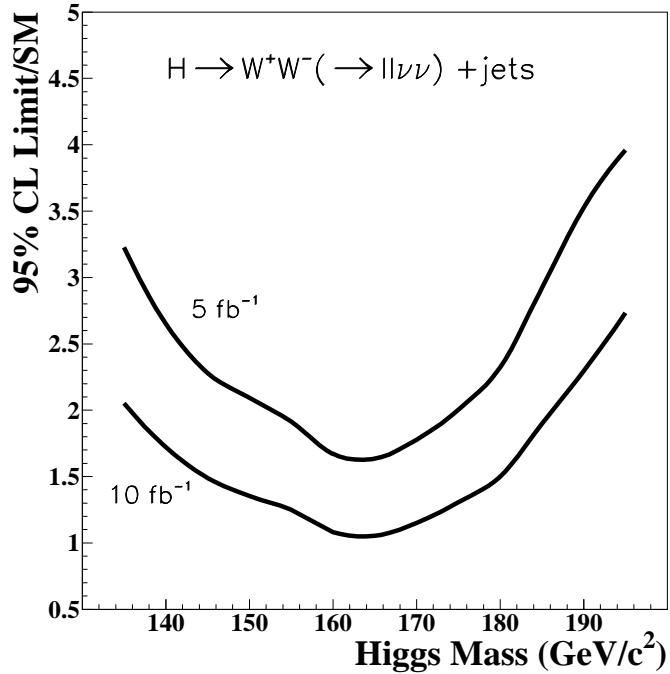,width=9cm}}}
\caption[]{Expected $95\,\%$ confidence level limit expressed in terms of the ratio of $\sigma\times\BR$ over the corresponding value in the Standard Model. Results are given as function of the Higgs mass  (in $\gev/c^2$) with $5\,$fb$^{-1}$ and $10\,$fb$^{-1}$ of integrated luminosity for two experiments combined. }
 \label{fig:results}
\end{figure}

Table~\ref{tab:signals} and Figure~\ref{fig:results} show the expected effective cross-sections (in fb) for the three signal processes and the three event selections considered in Sections~\ref{sec:selection1}-\ref{sec:selection3}. It is important to note that the event selections were optimized for $M_H=165\,\gev/c^2$ and no further mass dependent optimization was implemented. Therefore, when evaluating sensitivity  the background contribution remains unchanged for different Higgs mass hypotheses. A mass optimization is expected to yield additional sensitivity, especially for Higgs masses $M_H<150\,\gev/c^2$ where discriminants such as the scalar sum of the leptons and $\sla{p_T}$ of the transverse mass of the lepton-neutrino system yield separation against $WW$ and $t\overline{t}$ processes. For Higgs masses $M_H>170\,\gev/c^2$ the relaxation of the upper bound of the leptonic invariant mass can enhance the signal contribution while maintaining the signal to background ratio. 

\begin{table}[t]
\begin{center}
\begin{tabular}{||c||c|c|c|c|c|c|c|c|c|c|c|c|c||}
\hline 
$M_H$ & 135 &  140 &     145 &  150 &     155  &  160 &     165 &  170 &     175 &  180 &     185  &       190 &  195 \\  \hline  
 5fb$^{-1}$ & 3.2 & 2.7 & 2.3 & 2.1 & 1.9  & 1.7  &  1.6 &  1.8 & 2.0 & 2.3 & 2.9 & 3.5 & 4 \\ \hline
  10fb$^{-1}$ & 2.1 & 1.7 & 1.5 & 1.4 & 1.2  & 1.1  &  1.1 &  1.2 & 1.3 & 1.5 & 1.9 & 2.3 & 2.7 \\ \hline
\end{tabular}
\caption{Expected $95\,\%$ confidence level limit expressed in terms of the ratio of $\sigma\times\BR$ over the corresponding value in the Standard Model. Results are given as function of the Higgs mass  (in $\gev/c^2$) with $5\,$fb$^{-1}$ and $10\,$fb$^{-1}$ of integrated luminosity for two experiments combined. }
\label{tab:results}
\end{center}
\end{table}

The cuts used in the event selections presented in Sections~\ref{sec:selection1}-\ref{sec:selection3} are the result of the optimization of the $95\,\%$ confidence level limit for each channel individually. The event selections have not been optimized here to achieve the best limit for the three channels combined. 
The expected exclusion limit was calculated using a likelihood technique~\cite{ATL-PHYS-2003-008,physics_03_12050}. Table~\ref{tab:results} shows the expected $95\,\%$ confidence level limit as a function of the Higgs mass with 5\,fb$^{-1}$ and $10\,$fb$^{-1}$ of integrated luminosity (for two experiments combined).

In conclusion, the searches of a Higgs boson using $W^+W^-+jets$ with the event selections presented in this paper could further enhance the sensitivity of the Tevatron experiments reported in~\cite{CDFWWWinter07,D0WWWinter07_1,D0WWWinter07_2}.\footnote{The CDF and D0 analyses suppress events with large jet multiplicity either via the application of an explicit veto on events with multiple jets or the application of a cut on $H_T$.}
It is important to note that we use LO cross-sections for the $gg\rightarrow Hj$ process. The NLO K-factors for this process are expected to be large, thus significantly enhancing the sensitivity of analysis I. In addition, no multivariate techniques have been implemented. With tagging hadronic jets the complexity of the final state increases and with it the relative sensitivity of  a multivariate analysis with respect to the simple cut-based approach used here. In particular, variables like the transverse mass of the Higgs-leading jet system, the invariant mass of the two leading jets can be used as additional discriminating variables when appropriate. The analysis strategy presented here is conservative and leaves room for significant improvement in the sensitivity.

\section{Acknowledgements}

The authors are most grateful to Y.~Fang, L.~Flores-Castillo, T.~Han, M.~Herndon, F.~Petriello and T.~Vickey for most valuable  comments and suggestions. This work was supported in part by the United States Department of Energy through Grant No. DE-FG0295-ER40896.


\begin{mcbibliography}{10}

\bibitem{np_22_579}
S.~L.~Glashow,
\newblock  Nucl. Phys. {\bf B22}  (1961)~ 579\relax
\relax
\bibitem{prl_19_1264}
S.~Weinberg,
\newblock  Phys. Rev. Lett. {\bf 19}  (1967)~ 1264\relax
\relax
\bibitem{sal_1968_bis}
A.~Salam,
\newblock  Proceedings to the Eigth Nobel Symposium, May 1968, ed: N.~Svartholm
  (Wiley, 1968) 357\relax
\relax
\bibitem{pr_2_1285}
S.L.~Glashow, J.~Iliopoulos and L.~Maiani,
\newblock  Phys. Rev. {\bf D2}  (1970)~ 1285\relax
\relax
\bibitem{prl_13_321}
F.~Englert, R.~Brout,
\newblock  Phys. Rev. Lett. {\bf 13}  (1964)~ 321\relax
\relax
\bibitem{pl_12_132}
P.~W.~Higgs,
\newblock  Phys. Lett. {\bf 12}  (1964)~ 132\relax
\relax
\bibitem{prl_13_508}
P.~W.~Higgs,
\newblock  Phys. Rev. Lett. {\bf 13}  (1964)~ 508\relax
\relax
\bibitem{pr_145_1156}
P.~W.~Higgs,
\newblock  Phys. Rev. {\bf 145}  (1966)~ 1156\relax
\relax
\bibitem{prl_13_585}
G.~S.~Guralnik, C.R.~Hagen and T.W.B.~Kibble,
\newblock  Phys. Rev. Lett. {\bf 13}  (1964)~ 585\relax
\relax
\bibitem{pr_155_1554}
T.W.B.~Kibble,
\newblock  Phys. Rev. {\bf 155}  (1967)~ 1554\relax
\relax
\bibitem{prl_40_11_692}
H.M.~Georgi, S.L.~Glashow, M.E.~Machacek and D.V.~Nanopoulos,
\newblock  Phys. Rev. Lett. {\bf 40}  (1978)~ 11\relax
\relax
\bibitem{pl_136_196}
R.~Cahn and S.~Dawson,
\newblock  Phys. Lett. {\bf B136}  (1984)~ 196\relax
\relax
\bibitem{pl_148_367}
G.~Kane, W.~Repko and W.~Rolnick,
\newblock  Phys. Lett. {\bf B148}  (1984)~ 367\relax
\relax
\bibitem{prl_82_25}
T.~Han and R.-J.~Zhang,
\newblock  Phys. Rev. Lett. {\bf 82}  (1999)~ 24\relax
\relax
\bibitem{pr_59_093001}
T.~Han, A.S.~Turcot and R.-J.~Zhang,
\newblock  Phys. Rev. {\bf D59}  (1999)~ 093001\relax
\relax
\bibitem{JHEP_12_5}
D.L.~Rainwater and D.~Zeppenfeld,
\newblock  JHEP {\bf 12} (1997) 5\relax
\relax
\bibitem{pr_60_113004}
D.L.~Rainwater and D.~Zeppenfeld,
\newblock  Phys. Rev. {\bf D60}  (1999)~ 113004\relax
\relax
\bibitem{pr_59_014037}
K.~Hagiwara, D.L.~Rainwater and D.~Zeppenfeld,
\newblock  Phys. Rev. {\bf D59}  (1999)~ 014037\relax
\relax
\bibitem{pr_61_093005}
T.~Plehn, D.L.~Rainwater and D.~Zeppenfeld,
\newblock  Phys. Rev. {\bf D61}  (2000)~ 093005\relax
\relax
\bibitem{pl_431_410}
S.~Abdullin {\it et al.},
\newblock  Phys. Lett. {\bf B431}  (1998)~ 410\relax
\relax
\bibitem{pl_611_60}
B.~Mellado, W.~Quayle and Sau Lan Wu,
\newblock  Phys. Lett. {\bf B611}  (2005)~ 60\relax
\relax
\bibitem{EurPhysJ_32_19}
S.~Asai {\it et al.},
\newblock  Eur. Phys. J. C 32 (2004) s19-s54\relax
\relax
\bibitem{CMSPTDR}
CMS Collaboration,
\newblock  CMS PTDR V.2: Physics Performance,
\newblock  CERN/LHCC 2006-021\relax
\relax
\bibitem{cpc_82_74}
T.~Sj\"ostrand,
\newblock  Comp. Phys. Comm. {\bf 82} (1994) 74\relax
\relax
\bibitem{cpc_135_238}
T.~Sj\"ostrand {\it et al.},
\newblock  Comp. Phys. Comm. {\bf 135} (2000) 238\relax
\relax
\bibitem{JHEP_0101_010}
G.~Corcella {\it et al.},
\newblock  JHEP {\bf 0101} (2001) 010\relax
\relax
\bibitem{JHEP_0207_012}
J.~Pumplin {\it et al.},
\newblock  JHEP {\bf 0207} (2002) 012\relax
\relax
\bibitem{MCFM}
J.~Campbell and K.~Ellis,
\newblock  MCFM,
\newblock  http://mcfm.fnal.gov/\relax
\relax
\bibitem{JHEP_0307_001}
M.L.~Mangano {\it et al.},
\newblock  JHEP {\bf 0307} (2003) 001\relax
\relax
\bibitem{CDF8148}
The CDF Collaboration,
\newblock  Combination of CDF Top Quark Pair Production Cross Section
  Measurement with up to $760\,$pb$^{-1}$,
\newblock  CDF Note 8148\relax
\relax
\bibitem{JHEP_0206_029}
S.~Frixione and B.R.~Webber,
\newblock  JHEP {\bf 0206} (2002) 029\relax
\relax
\bibitem{JHEP_0308_007}
S.~Frixione and B.R.~Webber,
\newblock  JHEP {\bf 0308} (2003) 007\relax
\relax
\bibitem{CDFWWWinter07}
The CDF Collaboration,
\newblock  Search for $H\rightarrow WW^*$ Production with Matrix Element
  Methods in $pp$ Collisions at $\sqrt{s}=1.96\,\tev$,
\newblock  CDF Note 8774\relax
\relax
\bibitem{Herndon07}
M.~Herndon,
\newblock  private communication\relax
\relax
\bibitem{ATL-PHYS-2003-008}
K.~Cranmer, B.~Mellado, W.~Quayle and Sau Lan Wu,
\newblock  Confidence Level Calculations in the Search for Higgs Bosons Decay
  $H\rightarrow W^+W^- \rightarrow l^{+}l^{-}\sla{p_{T}}$ Using Vector Boson
  Fusion,
\newblock  ATLAS Note ATL-PHYS-2003-008 (2003)\relax
\relax
\bibitem{physics_03_12050}
K.~Cranmer, B.~Mellado, W.~Quayle and Sau Lan Wu,
\newblock  Challenges of Moving the LEP Higgs Statistics to the LHC,
\newblock  physics/0312050 (2003)\relax
\relax
\bibitem{D0WWWinter07_1}
The D0 Collaboration,
\newblock  Search for the Higgs Boson in $H\rightarrow WW^*\rightarrow\mu\mu$
  Decays with $930\,pb^{-1}$ at D0 in Run II,
\newblock  Conference Note 5194-CONF\relax
\relax
\bibitem{D0WWWinter07_2}
The D0 Collaboration,
\newblock  Search for the Higgs Boson in $H\rightarrow
  WW^*\rightarrow\mu\tau_{had}$ Signature with $1\,fb^{-1}$ at D0 in Run II,
\newblock  Conference Note 5332-CONF\relax
\relax
\end{mcbibliography}

\end{document}